\documentclass{tcibook}
\usepackage{fancyhea}
\usepackage{work}
\usepackage{bm}       %    enables bold math symbols  e.g.  \bm{\gamma}
\usepackage{graphicx}
\usepackage{hyperref}      % hypertext links %%ARXIV

\newcommand{\nc}{\newcommand}  

\def\beq{\begin{equation}}
\def\eeq#1{\label{#1}\end{equation}}
\def\eeqn{\end{equation}}

\newenvironment{Eqnarray}%
   {\arraycolsep 0.14em\begin{eqnarray}}{\end{eqnarray}}
\def\beqa{\begin{Eqnarray}}
\def\eeqa#1{\label{#1}\end{Eqnarray}}
\def\eeqan{\end{Eqnarray}}

\nc{\ra}{\rightarrow}  
\nc{\slsh}{\slash\hspace*{-0.22cm}}
\def\Re{{\cal R \mskip-4mu \lower.1ex \hbox{\it e}\,}}
\def\Im{{\cal I \mskip-5mu \lower.1ex \hbox{\it m}\,}}

\nc{\vev}[1]{ \left\langle {#1} \right\rangle }
\nc{\bra}[1]{ \langle {#1} | }
\nc{\ket}[1]{ | {#1} \rangle }
\nc{\fb}{\,{\rm fb}^{-1}}
\nc{\ev}{{\rm eV}}
\nc{\kev}{{\rm keV}}
\nc{\Mev}{{\rm MeV}}
\nc{\gev}{{\rm GeV}}
\nc{\tev}{{\rm TeV}}
\nc{\mev}{{\rm MeV}}

\def\del{\partial}
\def\Dslash{\not{\hbox{\kern-4pt $D$}}}
\def\dslash{\not{\hbox{\kern-2pt $\del$}}}
\def\pslash{\not{\hbox{\kern-2pt $p$}}}
\def\ETmiss{ \not{\hbox{\kern-4pt $E$}}_T }

\def\msb{{\bar{\ssstyle M \kern -1pt S}}}

\begin{document}

\def\bibname{References}
\bibliographystyle{plain}

\raggedbottom

\pagenumbering{roman}

\parindent=0pt
\parskip=8pt
\setlength{\evensidemargin}{0pt}
\setlength{\oddsidemargin}{0pt}
\setlength{\marginparsep}{0.0in}
\setlength{\marginparwidth}{0.0in}
\marginparpush=0pt

% The content begins here

\pagenumbering{arabic}

\renewcommand{\chapname}{chap:intro_}
\renewcommand{\chapterdir}{.}
\renewcommand{\arraystretch}{1.25}
\addtolength{\arraycolsep}{-3pt}

%%%%%%%%%%%%%%%%%%%%%%%%%%%%%%%%%%%%%%%%%%%%%%%%%%%
%%%%%%%%%%%%%%%%%%%%%%%%%%%%%%%%%%%%%%%%%%%%%%%%%%%
%%%     All of your files should be in a subdirectory.  Here the
%%%     subdirectory is called Magnetism  .   The title of your
%%%     report should be   wgreport.tex in that subdirectory.  Input
%%%     that file here
%%%%%%%%%%%%%%%%%%%%%%%%%%%%%%%%%%%%%%%%%%%%%%%%%%%%
%%%%%%%%%%%%%%%%%%%%%%%%%%%%%%%%%%%%%%%%%%%%%%%%%%%

%%%%%% Computing Chapter  %%%%%%%%%%%%%%%%
 
\chapter{Computing: Energy Frontier Sub-Group Report}
\label{chap:E2}

%%%%%%%%%%%%%%%%%%%%%%%%%%%%%%%%%%%%%%%%%%%%%%%%%%%%%%%%%%%
%%%%%%%%%%%%%%%%%%%%%%%%%%%%%%%%%%%%%%%%%%%%%%%%%%%%%%%%%%%
%%%%%%%%%%%%%%%%%%%%%%%%%%%%%%%%%%%%%%%%%%%%%%%%%%%%%%%%%%%
%%%%%%%%%%%%%%%%%%%%%%%%%%%%%%%%%%%%%%%%%%%%%%%%%%%%%%%%%%%
\begin{center}\begin{boldmath}

% list of HSPAW authors

%\hyphenpenalty 10000

\begin{center}

\begin{large} {\bf Conveners: I. Fisk (Fermilab), J. Shank (Boston University)} \end{large}

\end{center}

%\hyphenpenalty 1000

%Conveners are also listed separately in authorlist.tex

\end{boldmath}\end{center}

%%%%%%%%%%%%%%%%%%%%%%%%%%%%%%%%%%%%%%%%%%%%%%%%%%%%%%%%%%%
%%%%%%%%%%%%%%%%%%%%%%%%%%%%%%%%%%%%%%%%%%%%%%%%%%%%%%%%%%%
%%%%%%%%%%%%%%%%%%%%%%%%%%%%%%%%%%%%%%%%%%%%%%%%%%%%%%%%%%%
%%%%%%%%%%%%%%%%%%%%%%%%%%%%%%%%%%%%%%%%%%%%%%%%%%%%%%%%%%%

\section{Introduction}
\label{sec:comp-intro}

In an attempt to get a reasonable prediction of the magnitude of
changes that could be expected from a new program in the next ten
years, we look back on the changes between the Tevatron and LHC over
the last 10 years.  In 2003 the Tevatron was in the 3rd year of Run2
and comparing it to the third year of LHC in 2012 there is a rough
factor of 10 in the metrics associated with data acquisition and event
complexity as can be seen in Table~\ref{tab:compare_daq}. In addition
to the data acquired, the computing scales with the collaboration and
activity needs and those metrics are shown in
Table~\ref{tab:compare_coll}.

\begin{table}[t]
\begin{center}
\begin{tabular}{lll}
Metric & Tevatron (2003) & LHC (2012) \\ \hline
Trigger rate & 50 Hz & 500 Hz \\
Prompt reconstruction rate/week & 13 M Events & 120 M events \\
Rereconstruction rate & 100 M events per month & 800 M--1 B events per month \\
Reconstructed size & 200 KB & 1--2 MB \\
AOD size & 20 KB & 200--300 KB \\
Reconstruction time & 1--2 s on CPUs of the time & $\approx$ 10 s on CPUs of the time \\ \hline
\end{tabular}
\caption{Relevant data acquisition and complexity metrics comparing Tevatron and LHC at similar points.}
\label{tab:compare_daq}
\end{center}
\end{table}

\begin{table}[t]
\begin{center}
\begin{tabular}{lll}
Metric & Tevatron (2003) & LHC (2012) \\ \hline
Collaboration size & 800 & 2000--3000 \\
Number of individual analysis submitters per day & 100 & 300--400 \\
Number of total analysis submitters & 400 & $>$ 1000 \\ \hline
\end{tabular}
\caption{Relevant collaboration and participation metric comparing Tevatron and LHC at similar points.}
\label{tab:compare_coll}
\end{center}
\end{table}

Combining the increase in complexity, which is reflected in event size
and reconstruction time, with the increases in physics triggers and
the improvements in computing technology the total capacity increase
is much larger.  The globally distributed nature of LHC computing,
which includes the majority of the computing capacity located away
from the host lab, places much higher expectation on the networking
and data handling.  Both of these effects can be seen in
Table~\ref{tab:compare_comp}.

\begin{table}[t]
\begin{center}
\begin{tabular}{lll}
Metric & Tevatron (2003) & LHC (2012) \\ \hline
Remote computing capacity & 15 KHS06 (DZero Estimated) & 450 KHS06 (CMS) \\
User jobs launched per day & 10 K per day & 200--300 K jobs per day \\
Disk capacity per experiment in PB & 0.5 PB & 60 PB \\
Data on tape per experiment  & 400 TB & 70 PB \\
MC processing capacity per month & 3 M & 300 M \\
\qquad for full simulation &\\
Data served from dCache at FNAL per day & 25 TB & 10 PB \\
Wide area networking from host lab & 200 Mb/s & 20000 Mb/s \\
Inter VO transfer volume per day & 6 TB (DZero SAM) & 546 TB (ATLAS) \\ \hline
\end{tabular}
\caption{Relative capacity comparisons between Tevatron and LHC at similar points.}
\label{tab:compare_comp}
\end{center}
\end{table}

The processing capacity between the 2 programs has increased by a
factor of 30, which is almost exactly what would be expected by the 2
year Moore's law doubling cycle.  This is also reflected in the number
of user jobs, which will normally expand to fill the available
resources.  The disk capacity, the local data served, the wide area
networking from the host lab, and the inter-site transfers are all
increased by a factor of 100.  This is caused by the change in
computing model to have a much higher degree of distribution, but even
more impacted by the factor of 10 increase in trigger rate and the
factor of 10 increase in event size.  The full event simulation
capacity is also 100 times larger in the LHC program, which may
indicate the larger importance of simulation.  The factor of 30
increase in processing with a factor of 100 increase in IO and storage
makes an argument that processing scales as what can be accommodated
by Moore's law, and storage and IO scale with trigger rate and event
size.  The LHC has been successful even though these two numbers have
not increased at the same rate, but points toward the effort expended
in making efficient code and points toward issues facing computing for
the Energy Frontier moving forward.

\begin{figure}[htb]
\begin{center}
\includegraphics{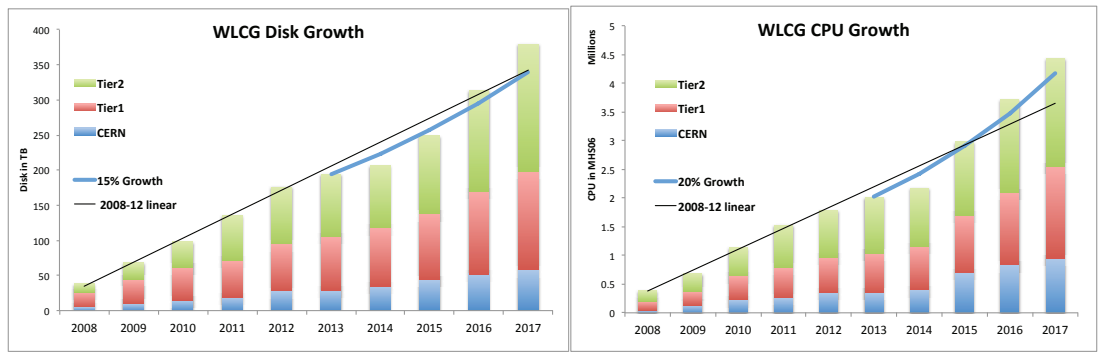}
\caption{The CPU and disk growth through the first 6 years of the LHC program
and projections to 2017.}
\label{fig:growth}
\end{center}
\end{figure}

The increase in LHC computing and disk storage is shown in
Figure~\ref{fig:growth}.  The CPU increases at a rate of 363 KHS06 per
year and the disk at 34 PB a year on average.  The roughly linear
increase is the combination of three separate periods that average to
linear.  The period 2008--2010 was the procurement ramp for
LHC as the scale of the available system was tested and commissioned.
The period 2010--2013 is the first run during which the computing and
storage increased at a rate determined by the need to process
and analyze the incoming data.  
The resources needed to accommodate the higher trigger
rate and event complexity expected in the second run define 2015.  The
three periods roughly average out to a linear increase.

In Energy Frontier computing data tends to be analyzed intensively at
the beginning and then archived and accessed less as most of the
relevant results are gleaned from the data in the first few years
after collection.  Therefore, the growth curves below do not scale
with total integrated luminosity but indicate that more computing is
needed per unit time as trigger rates and event complexity increase.
It is not reasonable to expect that the techniques currently used to
analysis data in the Energy Frontier will continue to scale
indefinitely.  The Energy Frontier will need to adopt new techniques
and methods moving forward.

%%%%%%%%%%%%%%%%%%%%%%%%%%%%%%%%%%%%%%%%%%%%%%%%%%%%%%%%%%%%%%%%%%%%%%%%%
%%
%%   use this format to include an .pdf figure into your paper
%%
%%%%%%%%%%%%%%%%%%%%%%%%%%%%%%%%%%%%%%%%%%%%%%%%%%%%%%%%%%%%%%%%%%%%%%%%%%%

If we extrapolate out 10 years, LHC computing would have roughly 3
times the computing expected in 2015, which is much lower than Moore's
law doubling expectations.  LHC would reach nearly 800 PB of disk by
2023, which is again roughly a factor of 3 over 2015.  The LHC numbers
are probably sustainable with these rates and budgets that are close to
flat for computing.  There are potential efficiency improvements and
new techniques that will be discussed below.  Of more concern is a
potential new program like a Super LHC, an ILC, or a LEP3 where the
luminosity or complexity increases dramatically.  Computing would then not
be on the curve in Figure~\ref{fig:growth}, but would be increased comparable
to the
difference between Tevatron Run2 and LHC Run1.  The proposed changes
below will potentially help make better use of the computing available
for LHC, and will be critical as Energy Frontier machines begin to
have data rates and conditions that would be expected in Intensity
Frontier computing.

\section{Trends to specialized systems and computing as a service}
\label{sec:comp-clouds}

With the move to Linux more than a decade and a half ago, Energy
Frontier computing has been relying on ever increasing capacity
provided by consistent and rather generic x86 hardware.  The vast
majority of applications are compiled with open source compilers and
not optimized for individual architectures.  It is understood that the
efficiency of the application for using the potential of the CPU is
low, but the volume of available computing is large and the
optimization has been to maintain the ability to run everywhere.
Before the move to Linux, experiments often supported several
platforms and compiled with a variety of dedicated compilers.  Moving
forward there are two trends that point in contradictory directions:
specialized hardware and computing as a service.

On one side, the Energy Frontier will need to evolve to use alternative
computing architectures and platforms because the focus of industry
development is moving into areas that are not the classic server CPU,
and there is the potential for dramatic increases in performance that
can change the slope of the curves in Figure~\ref{fig:growth} .  The
cost of this specialization is complexity and heterogeneity in the
computing system.  The machines that make up nearly all of the
processing resource capacity represent a small fraction of total
processing sales, and industry investments are in areas like GPUs, low
power applications for mobile computing, and specialized coprocessors.
All of these represent challenges and potential gains.  Using GPUs
introduces significant diversity to the system, complicates the
programming, and changes the approaches used in scientific
calculation, but can increase performance by orders of magnitude for
specific types of calculations.  Coprocessors have similar potential
improvement gains, increase in the diversity and complexity of the
system, and additional programming challenges.  Low power mobile
platforms are most interesting when they are combined into a massively
parallel, specialized system where a single box may have the same
number of cores as a remote computing center does today.  These
systems would be used more like a super computer and less like a batch
farm, which will require the field to grow expertise in this much more
interconnected computing environment.

Specialized hardware and architectures are likely to be deployed beginning
in extremely well controlled environments like trigger farms and other
dedicated centers where the hardware can be controlled and specified.
The next phase is likely to be scheduleable dedicated specialized
systems to permit large-scale calculations to achieve a goal similar
to making a super computer center request.  Large scale clusters of
specialized hardware owned by the experiment are likely to come last,
and are only likely to come if they can completely replace a class of
computing resources and perform a function at a reduced cost and
higher efficiency.

The other trend impacting Energy Frontier computing is the move to
computing as a service and other ``cloud'' solutions.  Currently,
commercial offerings, academic resources, and opportunistic resources
are all being offered through cloud provisioning techniques.  While
commercial solutions are still more expensive than well used dedicated
resources, there is a steady decrease in the pricing.  Academic clouds
function largely like other academic clusters but the cloud
environment expects the user to build up more of the services.
Opportunistic computing is an interesting growth area with a growing
number of resources with under utilized systems available,
particularly at night, being offered for applications that can make
effective use of limited duration or unpredictable duration computing.
We have seen a variety of places propose cloud-provisioning tools as
the primary interface to use the computing. While have not seen a site
contract with a commercial cloud provider to meet the obligations to
an experiment, it will come if the price continues to drop and the
experiments can make easy access of the resources through the
provisioning tools.  Small-scale clusters in expensive places, without
a history of computing, will likely be the first to be outsourced.

\section{Becoming more selective}
\label{sec:comp-select}

One trend that is visible from the Tevatron to the LHC is that while
the processing has increased largely with what would be expected from
Moore's law and relatively flat budgets, the storage requirements have
grown much faster.  The larger number of sites and the need for local
caches, the increase in trigger rates, and the larger event sizes
drives the need for storage.  For searches there is a case for storing
potentially interesting events and applying various hypotheses to look
for new physics.  For some searches and many measurements an approach
where much more of the processing and analysis is done with the
initial data collection and only synthesized output is archived has
the potential for preserving physics while reducing the offline
processing and storage needs.  Already the ALICE experiment is
proposing mostly on-line reconstruction after LHC Long Shutdown 2
(LS2).  In future accelerators like LEP3, a calibration run can
collect the entire LEP1 data set in 10 minutes.  There will be strong
motivations to reconstruct and calibrate on-line and write only
constants.

A change of the mentality that a higher trigger rate is always better,
and that any event selected should be protected from collection
through data preservation will be a change for the Energy Frontier
where the techniques used have been consistent through several
generations of machines.  As the Energy Frontier trigger rate
approaches numbers normally associated with the Intensity Frontier, the
techniques used in the Intensity Frontier will need to be considered.
In general, Energy Frontier experiments should expect to be more
selective and transform more of the reconstruction, calibration, and
analysis into a close to real time environment, if a sustainable
solution to computing is to be realized moving into the more intense
realms.

Similarly to how long and in what formats we store data, the same
issue exists for all derived data formats and simulation, in
particular.  A change for the LHC is the amount of simulation produced
compared to the number of events collected, which is often one-to-one
or more.  The generation and reconstruction of simulation is the
majority of the organized processing resources, but the most expensive
resource is the storage.  Simulation and reconstruction are entirely
derived data and can be reproduced and already many of the
intermediate steps are treated as transient.  A trend in the Energy
Frontier will be moving more of the analysis steps into the production
chain and only keeping the final output, with the knowledge it can be
re-derived.

\section{Data management}
\label{sec:comp-dm}

The disk space in the current generation of Energy Frontier
experiments scales with the number of events collected (trigger rate)
and the complexity (event size).  There is 100 times more disk space
during the 3rd year of LHC running compared to the third year of Run2.
In 2015 the big LHC experiments will deal with approximately 10B new
events combining data and simulation, but the analysis format of that
is 3PB so 10 copies/versions could be stored on Tier-2s centers.
Multiple versions and previous years data are analyzed, but it is
clear that many replicas can be hosted and the current model
preferentially places jobs where the data is physically hosted.

In an environment where hardware resources are specialized and
scheduled, it will be important to queue a large volume of data and
feed the specialized systems that can conceivably process and generate
output into a local cache quickly.  From a data management and data
transfer perspective, this is very similar to how clusters are
currently managed with data sets transferred in advance.  In a cloud
provisioned environment the concept of what is local data begins to
lose its relevance.  Cloud storage does not necessarily need to sit
near processors.  Cloud storage can be categorized by size and IO
capacity, but modern Energy Frontier applications can be optimized to
not lose significant efficiency even under high latency as long as the
bandwidth is high.  In cloud provisioned environments the storage
needed to feed the processors can sit long distances away provided the
bandwidth is sufficient, and even storage and processors in the same
resource provider may not be physically close or may move as different
capacity is provided.

To serve a cloud provisioned system the data management system begins
to look like a content delivery network (CDN), which is what Energy
Frontier computing should work to deploy.  The data federations being
pursued by the current generation of detectors are rudimentary content
delivery networks.  In general, data federations are currently intended
to serve a portion of the applications, in which the majority of the data
is served from local storage.  Dynamic replication and clean up as
well as predictive placement and network awareness are all needed to
enable the CDN to become the primary source of the bulk,
non-specialized computing.  Moving to a CDN for data management with
no expectation from the application for data locality simplifies the
use of cloud resources, of opportunistic resources, and of local user
controlled systems.  It reserves the current deterministic
pre-placement for specialized systems where the hardware is expected
to be too fast to be served over wide-area access, or so specialized
and precious that the risk of losing time because of loss of access to
the data could not be tolerated.

CDN systems have the potential to introduce enormous flexibility in
how data is accessed by a variety of computing systems, but it puts
demands on the networks.  Other CDN systems for video distribution are
among the largest users of the networks to residential homes in the
US, and one could reasonably expect these distribution systems would
be some of the largest users of research and education networks.  For
optimized IO currently 50--100 KB/s per core is needed for
reconstruction and more for analysis.  A 10 K core processing farm,
which in ten years will likely be the average for a Tier-2 center
could be served for reconstruction with a 10--20 Gb/s.  Sustaining 10 K
of analysis jobs will require a Tier-2 to have a 100 Gb/s link if
remote storage is the primarily source.  Sites providing
infrastructure to serve data to multiple sites will require multiple
100 Gb/s export links within a decade.  As part of a comprehensive data
management system, intermediate caches may be automatically populated
and used by local systems to enable local access.

Development in commercial computing has put a focus on delivering
content either through CDNs or through peer-to-peer systems.  In the
next decade, computing processing for the Energy Frontier will evolve
to be less deterministic, with more emphasis on cloud provisioned
resources, opportunistic computing, local computing, and volunteer
computing.  The data management system needs to evolve to be much less
deterministic as well in order to make efficient use of the diverse
landscape of resources.

\section{Activities needed}
\label{sec:comp-activities}
In the evolving landscape, we need to make development investments for
future growth to be able to improve the service and efficiency for
current programs and to be be able to support a new more challenging
machine.  The current distributed computing environment for the LHC
experiments has been developed and deployed for more than 10 years.
It relies on reasonably consistent sites with common interfaces for
processing and storage requests.  Moving forward, Energy Frontier
computing should expect a transition to more shared and opportunistic
resources provided through a variety of interfaces.  Effort is needed
to allow the community to make effective use of the diverse
environments and to perform resource provisioning across dedicated,
specialized, contributed, opportunistic, and purchased resources.

There is the potential of specialized hardware to dramatically
increase processing speed of Energy Frontier experiments.  Many-core
and massively multi-core have the same capacity in single boxes that
small scale clusters have today.  Specialized hardware in GPUs perform
particular calculations many factors faster than generic hardware if
programmed properly.  Programming skills for massively multi-core and
GPU programming need to be acquired and developed in the community.
This transition is more difficult than a change of language, it is a
change in how code is designed and the approach to problems.
Investment and expertise will be needed.

With the expected diversity of computing resources, Energy Frontier
computing needs to develop a data management system that can deal with
all of them.  A system is needed that handles the placement of the data and
allows the operations team and analysis users to concentrate more on
execution of work flows and less on placement and location of data.  
The development of a data intensive content delivery network
should not be unique to one experiment, and should even be applicable
to several scientific domains, but will require commitment and effort
to develop.

%%%%%%%%%%%%%%%%%%%%%%%%%%%%%%%%%%%%%%%%%%%%%%%%%%
%%%%%%%%%%%%%%%%%%%%%%%%%%%%%%%%%%%%%%%%%%%%%%%%%%
%%%   Your subdirectory (here Magnetism) should include
%%%    the files:
%%%           wgreport.tex
%%%           authorlist.tex
%%%         and all needed figures in pdf format
%%%%%%%%%%%%%%%%%%%%%%%%%%%%%%%%%%%%%%%%%%%%%%%%%%%%
%%%%%%%%%%%%%%%%%%%%%%%%%%%%%%%%%%%%%%%%%%%%%%%%%%%%

\end{document}